\documentclass{ws-ijmpa}
\def\beq{\begin{equation}}

\begin{document}

\markboth{Authors' Names}
{Instructions for Typing Manuscripts (Paper's Title)}

\catchline{}{}{}{}{}
%
%%%%%%%%%%%%%%%%%%%%%%%%%%%%%%%%%%%%%%%%%%%%%%%%%%%%%%%%%%%%%%%%%%%%

\title{Chiral Dynamics  and Dubna-Mainz-Taipei Dynamical Model \\for
    Pion-Photoproduction Reaction
\\ }

\author{SHIN NAN YANG}

\address{Department of Physics and Center for Theoretical Sciences, National Taiwan University, \\
Taipei, 10617,
Taiwan \\
snyang@phys.ntu.edu.tw}

\maketitle

%\begin{history}
%\received{Day Month Year}
%\revised{Day Month Year}
%\end{history}

\begin{abstract}
We demonstrate that the Dubna-Mainz-Taipei (DMT) meson-exchange
dynamical model, which starts from an effective chiral Lagrangian,
for pion photoproduction provides an excellent and economic
framework to describe both the $\pi^0$ threshold production and the
$\Delta$ deformation, two features dictated by  chiral dynamics.

\end{abstract}

\ccode{PACS numbers: 11.30.Rd, 13.60.Le, 14.20.-c, 14.20.Gk}

An important feature of the low-energy QCD is the chiral symmetry.
Chiral symmetry is expected to show up in the parity doubling of
all hadronic states (Winer-Weyl mode), e.g., the proton with $J^P = 1/2^+$ would have
a $1/2^-$ partner . This is not observed  experimentally. Instead
the symmetry is broken spontaneously (Nambu-Goldstone mode)
which leads to the appearance of massless pseudoscalar
mesons. The opposite parity partner of the proton is a proton
plus a "massless pion".

Spontaneous chiral symmetry breaking (SCSB) has led to the
development of chiral perturbation theory (ChPT),  a low-energy
effective field theory of QCD.  It utilizes the concept of SCSB and
replaces the quark and gluon fields  by a set of fields  describing
the degrees of freedom of the observed hadrons. There is generally
good agreement between the ChPT predictions and
experiments\cite{Bernstein07}, including the $\pi^0$ photoproduction
near threshold where very precise measurements have been performed
and the ChPT calculation to one loop  $O(p^4)$ has been carried out
in the heavy-baryon formulation\cite{Bernard}.

The fact that opposite parity partner of the proton is a proton plus
a pion leads to the consequence that the $\pi N$ interaction in
momentum space takes the form $V_{\pi N}=g_{\pi N} \vec
\sigma\cdot\vec q$, where $\vec \sigma$ and $\vec q$ are the nucleon
spin and pion momentum, respectively. This strong $p$-wave $\pi N$
interaction gives rise to the $\Delta$ resonance and its
deformation\cite{Bernstein07a}, which has been observed in pion
photoproduction.

In this contribution, I will present a meson-exchange dynamical
model for pion photoproduction we recently developed in a
collaboration between groups at Dubna, Mainz, and Taipei
(DMT)\cite{KY99} which can describe well the pion-photoproduction
data from threshold to the first resonance region, including the
$\pi^0$ threshold production and $\Delta$ deformation. The DMT
dynamical model also starts from an effective chiral Lagrangian. The
effective Lagrangian is then used to construct a potential for use
in the scattering equation. The solutions of the scattering equation
will include rescattering effects to all orders and thereby
unitarity is ensured.

In a dynamical model for  pion photoproduction\cite{yang85}, the
$t$-matrix is given as $
t_{\gamma\pi}(E)=v_{\gamma\pi}+v_{\gamma\pi}g_0(E)\,t_{\pi N}(E),$
where $v_{\gamma\pi}$ is the $\gamma\pi$ transition potential, $g_0$
and $t_{\pi N}$ are the $\pi N$ free propagator and $t$ matrix,
respectively, and $E$ is the total energy in the c.m. frame.
Physical multipole amplitude in channel $\alpha$   then reads
as\cite{yang85}
\begin{eqnarray}
\hspace{-0.2cm}  t^{(\alpha)}_{\gamma\pi}(q_{E}, k_E;E+i\varepsilon) =
e^{i\delta_{\alpha}}\cos\delta_{\alpha}\,[ v^{(\alpha)}_{\gamma\pi}
  +   P\int^{\infty}_{0}d q'\frac{q'^2
R^{(\alpha)}_{\pi N}
(q_{E},q';E)\,v^{(\alpha)}_{\gamma\pi}(q',k_E)}{E-E_{\pi N}(q')}],
\end{eqnarray}
where $\delta_{\alpha}$, $ R_{\pi N}^{(\alpha)}$, $E_{\pi N}(q)$ and
$P$ denote the $\pi N$ phase shift, reaction matrix in channel
$\alpha$, total CM energy of momentum $q$, and principal value
integral, respectively; $k_E=\mid{\bf k}\mid$ is the photon momentum
and $q_E$ the pion on-shell momentum. The amplitudes $t_{\pi N}$ are obtained in a
meson-exchange $\pi N$ model\cite{hung} constructed in the
Bethe-Salpeter formalism and solved within Cooper-Jennings reduction
scheme. At low energies where resonances play little role, only
background part, $v^B_{\pi N}$ and $v^B_{\gamma\pi}$, which are
derived from an effective Lagrangian containing Born terms and
$\rho$ and $\omega$ exchanges in the $t$ channel, contribute.

For $\pi^0$ photoproduction from proton, we calculate the multipole
$E_{0+}$ near threshold by solving the following coupled channels
equation within a basis with physical pion and nucleon masses. It
leads to the following expression in $\pi^0 p$ channel:
\begin{eqnarray}
t_{\gamma\pi^0}(E)&= & v_{\gamma\pi^0}(E)+v_{\gamma\pi^0}(E)\,
g_{\pi^0 p}(E)\,t_{\pi^0 p\rightarrow \pi^0 p}(E) \nonumber\\& +&
v_{\gamma\pi^+}(E)\, g_{\pi^+ n}(E)\,t_{\pi^+ n\rightarrow \pi^0
p}(E)\,. \label{eq:coupled}
\end{eqnarray}
The $\pi N$ $t$-matrices are obtained by solving the coupled
channels equation for $\pi N$ scattering using the meson-exchange
model\cite{hung}. In Fig. 1, the prediction of DMT model for
$Re\,E_{0+}$ obtained without and with isospin symmetry assumption,
are shown in dashed and solid curves, respectively, and compared
with heavy-baryon ChPT results (dash-dotted curve)\cite{Bernard}.
Agreement of DMT prediction with the data
\begin{figure}[h]
\begin{flushright}
\begin{minipage}[t]{55mm}
%\centerline{\psfig{file=thresholdE0.eps,width=5.0cm}} \vspace*{8pt}
\centerline{\psfig{file=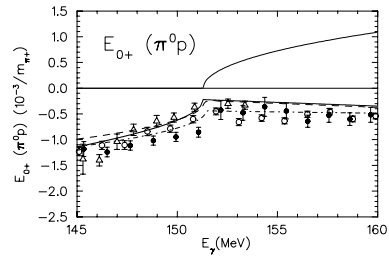,width=5.0cm}} \vspace*{8pt}
\end{minipage}
\end{flushright}
\begin{minipage}[t]{60mm}
\vspace{-4.00cm} \caption{Real  part of the $E_{0+}$ multipole for
$\gamma p\rightarrow \pi^0 p$. The dashed and solid curves are the
full DMT results obtained without and with isospin symmetry
assumption, respectively. The dash-dotted curve is the result of
ChPT. Data points  are from various
experiments\protect\cite{PLB01}.}
\end{minipage}
\end{figure}
\noindent

%\noindent
\newpage

%%%%%%%%%%%%%%%%%%%%%%%%%%%%%%%%%%%%%%%%%%%%%%%%%%%%%%%%%%%%%%%%%%%
\begin{figure}[h]
\begin{flushright}
\begin{minipage}[t]{55mm}
%\centerline{\psfig{file=Sigma.eps,width=5.0cm}} \vspace*{8pt}
\centerline{\psfig{file=figure2.eps,width=5.0cm}} \vspace*{8pt}
%\end{minipage}
%\end{flushright}
%\begin{minipage}[t]{60mm}
%\vspace*{-3.75cm}
\caption{Photon asymmetry at 159.5 MeV. The solid and dashed curves
are results of DMT and ChPT calculations. Data are from
Mainz\protect\cite{Schmidt01}.}
%\end{minipage}
%\end{flushright}
%\end{figure}
\vspace{0.5cm}
%\begin{figure}[pb]
%\begin{flushright}
%\begin{minipage}[t]{55mm}
%\centerline{\psfig{file=ME33-multipoles.eps,width=5.0cm}} 
\centerline{\psfig{file=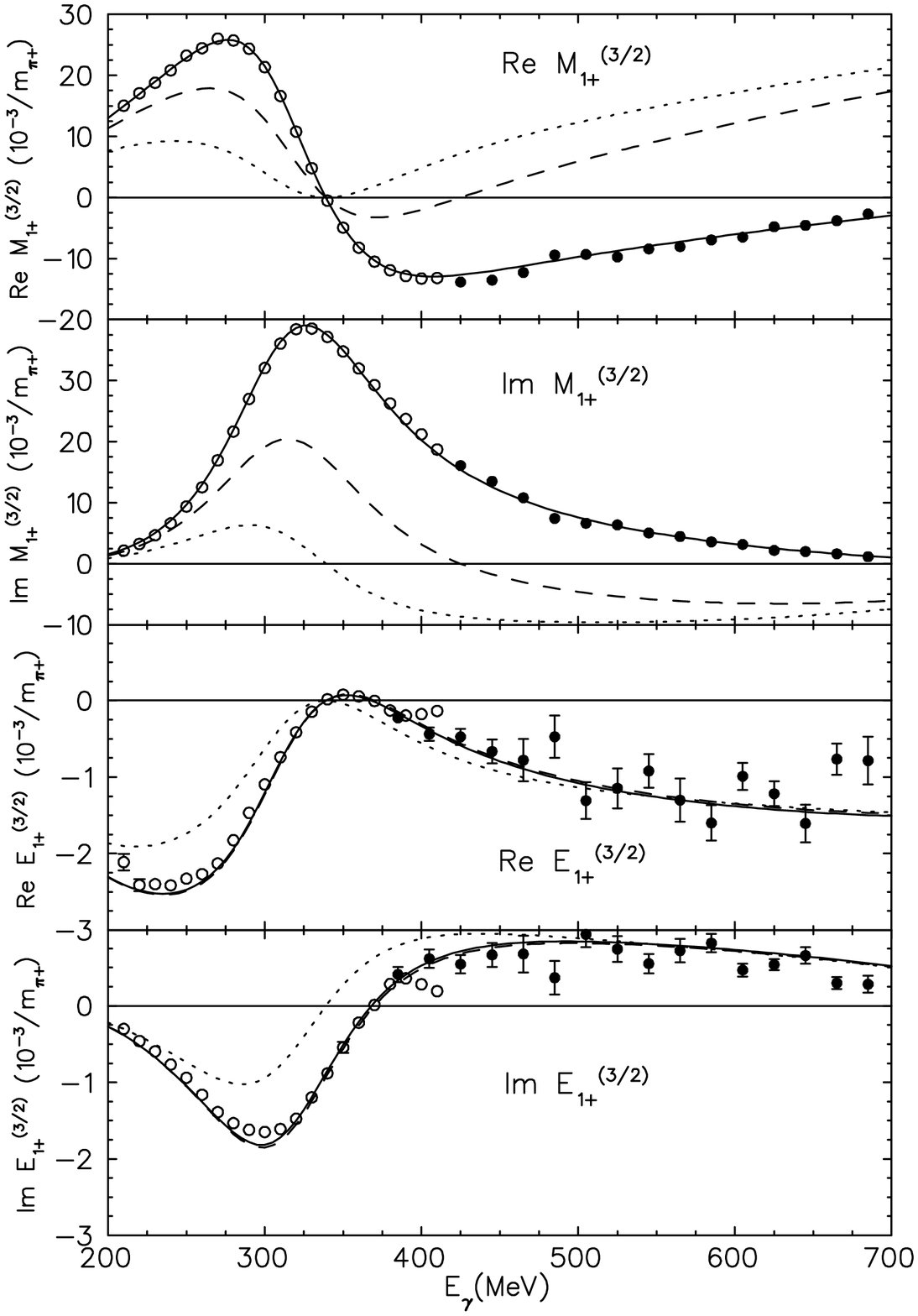,width=5.0cm}}
\caption{
Real and imaginary parts of the $M_{1+}^{(3/2)}$ and
$E_{1+}^{(3/2)}$ multipoles. Dotted and dashed curves are the
results for the $t_{\gamma\pi}^B$ obtained without and with
principal value integral contribution in Eq. (1), respectively.
Solid curves are the full results with bare $\Delta$ excitation. For
the $E_{1+}$ dashed and solid curves are practically the same due to
a very small value of the bare $G_{E2}$. The open and full circles
are the results from the analyses of Mainz\protect\cite{HDT} and
VPI\protect\cite{VPI97}. }
\end{minipage}
\end{flushright}
\end{figure}
\noindent
\begin{minipage}[t]{60mm}

\vspace{-19.0cm}

\noindent
and ChPT results are excellent.

\hspace{0.5cm} The
polarized linear photon asymmetry $\Sigma$ has been found to be very
sensitive to small $p$-wave multipoles\cite{PLB01}. DMT model (solid
curve) is not able to reproduce the data at $159.5$ MeV, of Ref.
\refcite{Schmidt01}, as shown in Fig. 2, while ChPT calculation of
$O(p^4)$ with six low-energy constants (dashed curve) is seen to be
able to describe the experiment reasonably well. However,
preliminary analysis of a new measurement at Mainz\cite{Hornidge10}
seems to agree with DMT's prediction.

\hspace{0.5cm} We now turn to the issue of the $\Delta(1232)$ deformation. In a
symmetric SU(6) quark model, $\Delta$ is in $S$ state and spherical.
The photo-excitation of the $\Delta$ could then proceed only via
$M1$ transition. The existence of a $D$ state in the $\Delta$ has
the consequence that the $\Delta$ is deformed and the photon can
excite a nucleon through electric $E2$ quadrupole transition.  In
pion photoproduction, $E2$  excitation  would give rise to
nonvanishing $E_{1+}^{(3/2)}$   multipole amplitude. Recent
experiments give $R_{EM} =
E_{1+}^{(3/2)}/M_{1+}^{(3/2)}=-(2.5\pm0.5)\%$\cite{Pascalutsa07}, a
clear indication of $\Delta$ deformation.

\hspace{0.5cm} In the (3,3) channel where $\Delta$ excitation plays an important
role, the transition potential $v_{\gamma\pi}$ consists of two terms
\begin{eqnarray}
v_{\gamma\pi}(E)=v_{\gamma\pi}^B + v_{\gamma\pi}^{\Delta}(E)\,,
\label{eq:tranpot}
\end{eqnarray}
where  second term of Eq. (\ref{eq:tranpot}) corresponds to the
contribution of bare $\Delta$, namely,  $\gamma N \rightarrow \Delta
\rightarrow \pi N$. We may then write

\vspace{-0.5cm}

\begin{eqnarray}
t_{\gamma\pi}= t_{\gamma\pi}^B + t_{\gamma\pi}^{\Delta}\,,
\label{eq:decomp}
\end{eqnarray}
\end{minipage}

\vspace{-.8cm}

\noindent where
$ t_{\gamma\pi}^B(E)=v_{\gamma\pi}^B+v_{\gamma\pi}^B\,g_0(E)\,t_{\pi
N}(E)\,\,$ and
$\,\,t_{\gamma\pi}^\Delta(E)=v_{\gamma\pi}^\Delta+v_{\gamma\pi}^\Delta\,
g_0(E)\,t_{\pi N}(E)$.

  By combining the
contributions of $t_{\gamma\pi}^B$ and $t_{\gamma\pi}^\Delta$ and
using the bare $\gamma N\Delta$ coupling constants $G_{M1}$ and
$G_{E2}$ for $M1$ and $E2$ transitions as free parameters, results
of our best fit to the resonant multipoles $M_{1+}^{(3/2)}$ and
$E_{1+}^{(3/2)}$ obtained in the analyses  of Mainz\cite{HDT} and
VPI group\cite{VPI97} are shown in Fig. 3 by solid curves. The
dashed curves denote the contribution from $t_{\gamma\pi}^B$ only.
The dotted curves represented the $K$-matrix approximation to
$t_{\gamma\pi}^B$, namely, without the principal value integral term
of Eq. (1) included.

For $M_{1+}^{(3/2)}$, one sees a large effect of the pion off-shell
rescattering (difference between dotted and dashed curves), which
results from the principal value integral part of  Eq. (1).
The total pion rescattering (dashed curves) contributes
for half of the $M_{1+}^{(3/2)}$ as seen in
Fig. 3 while the remaining half originates from the
bare $\gamma N \Delta$ excitation.
Furthermore, one sees that almost all of the $E2$ strength is generated
by the $\pi N$ rescattering.

At the resonance position $t_{\gamma\pi}^B$ vanishes within $K$-matrix
approximation and only principal value integral term survives. The
latter corresponds to the contribution where $\Delta$ is excited by
the pion produced via $v_{\gamma\pi}^B$. Consequently the addition
of this contribution to $t_{\gamma\pi}^{\Delta}$ can be considered
as a dressing of the $\gamma N \Delta$ vertex.  For
$E_{1+}^{(3/2)}$, the dominance of background and pion rescattering
contributions  leads to a very small bare value for electric
transition. We hence conclude that bare $\Delta$ is almost spherical
and the deformation observed experimentally arises mostly from the
long-range effect of the pion cloud, a manifestation of chiral
dynamics.

In summary, we have demonstrated that the DMT meson-exchange
dynamical model for pion photoproduction, which starts from an
chiral effective Lagrangian, provides an excellent and economical
framework in describing threshold $\pi^0$ production and $\Delta$
deformation, two key consequences of chiral dynamics.

\section*{Acknowledgments}
Results presented here are obtained in collaborations with G. Y.
Chen, D. Drechsel, S. S. Kamalov, and L. Tiator. This work is
supported in part by  the National Science Council of ROC under
grant NSC 98-2112-M002-006.

%\begin{thebibliography}{000} %for 3 digits
%\begin{thebibliography}{00}  %for 2 digits

\end{document}